\documentclass[aps,prd,onecolumn,floatfix,nofootinbib,superscriptaddress]{revtex4-1}
\usepackage[colorlinks,linkcolor=blue,anchorcolor=blue,citecolor=blue,urlcolor=blue,breaklinks=true]{hyperref}
\usepackage[libertine]{newtxmath}
\usepackage{graphicx}
\usepackage{slashed}
\usepackage{amsfonts}
\usepackage{amsmath}
\usepackage{blindtext}
\usepackage{microtype}
\usepackage{easyReview}
\usepackage{booktabs}
\usepackage{cleveref}
\usepackage{bm}
\usepackage{extarrows}

\newcommand{\vect}[1]{\boldsymbol{#1}}

\graphicspath{{.}{figures/}}

\def\ua{\uparrow}
\def\da{\downarrow}
\def\para{\parallel}

\begin{document}

\title{Nonperturbative photon $q\bar{q}$ light front wave functions from a contact interaction model}

\author{Chao Shi}
\email[]{cshi@nuaa.edu.cn}
\affiliation{Department of Nuclear Science and Technology, Nanjing University of Aeronautics and Astronautics, Nanjing 210016, China}

\author{Zhongtian Yang}
\affiliation{Department of Nuclear Science and Technology, Nanjing University of Aeronautics and Astronautics, Nanjing 210016, China}

\author{Xurong Chen}
\affiliation{Institute of Modern Physics, Chinese Academy of Sciences, Lanzhou 730000, China}

\author{Wenbao Jia}
\affiliation{Department of Nuclear Science and Technology, Nanjing University of Aeronautics and Astronautics, Nanjing 210016, China}

\author{Cuibai Luo}
\email[]{cuibailuo@ahnu.edu.cn}
\affiliation{Department of Physics, Anhui Normal University, Anhui 241002, China}

\author{Wenchang Xiang}
\email[]{wxiangphy@qq.com}
\affiliation{Physics Division, Guangzhou Maritime University, Guangzhou 510725, China}

\begin{abstract}

We propose a method to calculate the $q\bar{q}$ light front wave functions (LFWFs) of photon at low-virtuality, i.e., the light front amplitude of $\gamma^*\rightarrow q\bar{q}$ at low $Q^2$, based on a light front projection approach. We exemplify this method using a contact interaction model within  Dyson-Schwinger equations formalism and obtain the nonperturbative photon $q\bar{q}$ LFWFs. In this case, we find the nonperturbative effects are encoded in the enhanced quark mass and a dressing function of covariant quark-photon vertex, as compared to the leading order quantum electrodynamics photon $q\bar{q}$ LFWFs. We then use nonperturbative-effect modified photon $q\bar{q}$ LFWFs to study the inclusive deep inelastic scattering HERA data in the framework of the color dipole model. The results demonstrate that the theoretical description of data at low $Q^2$ can be significantly improved once the nonperturbative corrections are included in the photon LFWFs.

\end{abstract}
\maketitle

\section{INTRODUCTION\label{intro}}
The photon provides an important probe to hadron's internal structure, which historically helped found the Quantum Chromodynamics (QCD) \cite{Breidenbach:1969kd,Bloom:1969kc}. The photon is considered as a clean probe, as its interactions with quarks are primarily through Quantum Electrodynamics (QED). While this is true for photon with high virtuality $Q^2$, the situation gets more complicated at low $Q^2$. The reason is that although photon is an elementary particle, it can quantum fluctuate into other states with same quantum number in scattering processes, including $\gamma^*\rightarrow q\bar{q}$ that contains colored objects. These colored objects interact strongly at low scale and bring nonperturbative QCD effects. 

In a light-front (or light-cone) frame, the splitting amplitudes of $\gamma^*\rightarrow q\bar{q}$  are referred to as the photon $q\bar{q}$ light front wave functions (LFWFs) \cite{Dosch:1996ss,Forshaw:2003ki,Lappi:2020ufv}. They are interesting by themselves as they encode  $q\bar{q}$ components of photon. Meanwhile, the photon $q\bar{q}$ LFWFs are important input in the color dipole model,  which had been applied to the study of various processes such as inclusive and diffractive deep inelastic scattering (DIS) \cite{Nikolaev:1990ja, Kowalski:2003hm,Marquet:2007nf,Rezaeian:2012ji,Mantysaari:2016ykx,Mantysaari:2018nng}, diffractive vector meson production \cite{Kowalski:2006hc,Forshaw:2012im,ArroyoGarcia:2019cfl, Li:2021cwv} and ultra-peripheral heavy ion collisions \cite{Goncalves:2005yr,Lappi:2013am,Xie:2016ino,Goncalves:2017wgg}. For a few decades, photon LFWFs at leading order QED \cite{Dosch:1996ss} are employed in the color dipole model studies, until next leading order corrections from QCD became available in recent years \cite{Beuf:2016wdz,Beuf:2021qqa,Beuf:2022ndu,Beuf:2022kyp,Beuf:2021srj}.  
Meanwhile, the nonperturbative photon LFWFs were also considered in  literatures \cite{Forshaw:2003ki,Flensburg:2008ag,Berger:2012wx,Goncalves:2020cir}. Therein a model function is introduced to mimic the nonperturbative effect, with model parameters determined by phenomenological fitting of experiment data. It was found that the inclusion of the nonperturbative QCD corrections in the photon LFWFs improve the description of the experimental data for the observables that are sensitive to color dipoles of large size, namely with low $Q^2$ \cite{Berger:2012wx,Goncalves:2020cir}. It is also anticipated that the real photons ($Q^2=0$) produced in deeply virtual Compton scattering (DVCS) are strongly affected by the nonperturbative effects \cite{Flensburg:2008ag}.

On the other hand, to our best knowledge, a theoretical calculation on photon's nonperturbative $q\bar{q}$ LFWFs at low virtuality is absent in the literature to date. In this paper we will tackle this problem using a light-front projection method we introduced in \cite{Shi:2021taf}. Therein a projection formula to extract vector meson $q\bar{q}$ LFWFs from their covariant Bethe-Salpeter (BS) wave functions was given. As photon is not a composite particle, the meson BS wave functions should be replaced by the photon's  inhomogeneous BS wave function, which is essentially the covariant (un-amputated) quark-photon vertex. The nonperturbative effects in quark-photon vertex at low $Q^2$ had long been investigated by Dyson-Schwinger equations (DSEs) approach \cite{Maris:1999bh}. The authors found that by properly dressing the quark-photon vertex, the pion charge radius increases by 50\% and gets close to experimental value. Naturally, these nonperturbative effects can be conveyed to photon $q\bar{q}$ LFWFs. 

In this paper, we will explore the nonperturbative effects in photon $q\bar{q}$ LFWFs starting with a simplified model, i.e., the contact interaction model within the DSEs.  The contact interaction model had been widely employed in the study of various hadron properties \cite{Gutierrez-Guerrero:2010waf,Chen:2012txa,Xu:2015kta,Yin:2019bxe,Yin:2021uom,Xing:2022jtt,Zamora:2023fgl}. Despite the simplicity of the model, it provides a good ground to depict our projection method and calculation techniques. It also renders analytical results on photon $q\bar{q}$ LFWFs, showing intuitively how the nonperturbative effects are encoded in the  photon $q\bar{q}$ LFWFs through dressed scalar functions.

This paper is organized as follows. In section \ref{sec:LF-LFWF} we introduce the general formalism of photon $q\bar{q}$ LFWFs, as well as the light front projection formula to extract them from covariant quark-photon vertex. In section \ref{sec:CI}, we recapitulate the contact interaction model within the Dyson-Schwinger equations approach, and demonstrate the calculation of nonperturbative photon $q\bar{q}$ LFWFs with detail.  In section \ref{sec:DIS}, we show that the inclusion of nonperturbative photon LFWFs can improve the agreement between color dipole model calculations and small-$x$ inclusive DIS data at low $Q^2$. Finally we summarize in section \ref{sec:sum}. 

\section{Photon $q\bar{q}$ LFWFs \label{sec:LF-LFWF}}
Consider a virtual photon $\gamma^*$, including the real photon as the limiting case $Q^2\rightarrow 0$, which schematically has a Fock-state expansion on the light front as
\begin{align}
|\gamma^*_{\rm phys} \rangle=|\gamma^*_{\rm bare}\rangle+|e^+e^-\rangle_{\gamma^*}+\sum_{f=u,d,s...}|q_f\bar{q}_f\rangle_{\gamma^*}+...
\end{align}
Here the $|e^+e^-\rangle_{\gamma^*}$  can be calculated with perturbative QED. The $|q_f\bar{q}_f\rangle_{\gamma^*}$, on the other hand, is more complicated. At high virtuality, asymptotic freedom allows a perturbative calculation. Yet at low virtuality, the quark and antiquark components interact strongly and the system can be nonperturbative, where nonperturbative QCD method is called for. A general decomposition of $|q_f\bar{q}_f\rangle_{\gamma^*}$ reads
\begin{align}\label{eq:LFWF1}
	|q_f\bar{q}_f\rangle^\Lambda_{\gamma^*} &= \sum_{\lambda,\lambda';i,j}\int \frac{d^2 \vect{k}_T}{(2\pi)^3}\,\frac{dx}{2\sqrt{x\bar{x}}}\, \frac{\delta_{ij}}{\sqrt{3}} \Phi^{\Lambda,(f)}_{\lambda,\lambda'}(x,\vect{k}_T)\, b^\dagger_{f,\lambda,i}(x,\vect{k}_T)\, d_{f,\lambda',j}^\dagger(\bar{x},\bar{\vect{k}}_T)|0\rangle.
\end{align}
The $\Phi^{\Lambda,(f)}_{\lambda,\lambda'}$ is the $q\bar{q}$ LFWF of photon with helicity $\Lambda$ and quark (antiquark) of flavor $f (\bar{f})$ and spin $\lambda$ ($ \lambda'$). The $\Lambda=0, \pm 1$ and $\lambda=\ua$ or $\da$, denoted as $\ua=+$ and $\da=-$ for abbreviation in following. The $b^+$ and $d^+$ are creation operators of quark and antiquark, respectively. The $i$ and $j$  are the color indices. The $\vect{k}_T=(k^x,k^y)$ is the transverse momentum of the quark, and $\bar{\vect{k}}_T=-\vect{k}_T$ for antiquark \footnote{We take a frame where the virtual photon has no transverse momentum.}. The longitudinal momentum fraction carried by quark is $x=k^+/P^+$, with $\bar{x}=1-x$ for antiquark. Note we take the light-cone four vector convention as $A^{\pm} = \tfrac{1}{\sqrt{2}}(A^0 \pm A^3)$ and $\vect{A}_T=(A^1, A^2)$ throughout this paper. 

In  \cite{Shi:2021taf}, we have introduced the light front projection method to obtain $q\bar{q}$ LFWFs of vector mesons from their BS wave functions. The projection formula applies to real and virtual photon as well, e.g.,
\begin{align}\label{eq:chi2phi}
	\Phi^{\Lambda,(f)}_{\lambda,\lambda'}(x,\vect{k}_T)&=-\frac{1}{2\sqrt{3}}\int \frac{dk^- dk^+}{2 \pi} \delta(x Q^+-k^+) \nonumber\\
	&\hspace{5mm}\times\textrm{Tr}\left \{\Gamma_{\lambda,\lambda'}\gamma^+ S_f(k)[e_f e \Gamma^{\gamma^*,(f)}(k;Q)\cdot \epsilon_\Lambda(Q)]S_f(k-Q)  \right\}.
\end{align}
The $S(k)$ and $\Gamma^{\gamma^*}_\mu(k;Q)$ are the fully dressed quark propagator and (amputated) quark--photon vertex in the momentum space. They can be obtained by solving the quark gap equation and inhomogeneous BS equation, which will be addressed in a later section.  The $\epsilon^\mu_\Lambda(Q)$ is the  photon polarization vector. The $\Gamma_{\pm,\mp}=I\pm \gamma_5$ and $\Gamma_{\pm,\pm}=\mp(\gamma^1\mp i\gamma^2)$ project out corresponding quark-antiquark helicity configurations. Note that there is an implicit unit matrix in color space attached to quark propagators and vertices. The trace is taken over Dirac and color indices. 

Analogous to the case of vector meson, due to symmetry constraints, the $\Phi^{\Lambda,(f)}_{\lambda,\lambda'}(x,\vect{k}_T)$'s can generally be expressed with five independent scalar amplitudes $\psi(x,\vect{k}_T^2)$'s \cite{Carbonell:1998rj,Ji:2003fw,Shi:2021taf}, i.e.,
\begin{align}
	\hspace{00mm}\Phi_{\pm,\mp}^{0,(f)}&=\psi^{0,(f)}_{(1)},  \ \ \ \ \ 
	&\Phi_{\pm,\pm}^{0,(f)}&=\pm k_T^{(\mp)} \psi^{0,(f)}_{(2)}, \label{eq:phi1}\\
	\Phi_{\pm,\pm}^{\pm 1,(f)}&=\psi^{1,(f)}_{(1)},
	&\Phi_{\pm,\mp}^{\pm 1,(f)}&=\pm  k_T^{(\pm)}\psi^{1,(f)}_{(2)}, \notag \\
	\Phi_{\mp,\pm}^{\pm 1,(f)}&=\pm k_T^{(\pm)}\psi^{1,(f)}_{(3)},
	&\Phi_{\mp,\mp}^{\pm 1,(f)}&=(k_T^{(\pm)})^2\psi^{1,(f)}_{(4)}. \label{eq:phi2}
\end{align}
with $k_T^{(\pm)}=k^x \pm i k^y$, and
\begin{align}\label{eq:psi2}
	\psi^{1,(f)}_{(2)}(x,\vect{k}_T^2)&=-\psi^{1,(f)}_{(3)}(1-x,\vect{k}_T^2).
\end{align}
The convention is to take $+$ and $-$ signs in the same row of one equation at once, e.g. $\Phi_{\pm,\mp}^{\pm 1,(f)}=\pm  k_T^{(\pm)}\psi^{1,(f)}_{(2)}$ means $\Phi_{+,-}^{+ 1,(f)}=+  k_T^{(+)}\psi^{1,(f)}_{(2)}$ and $\Phi_{-,+}^{- 1,(f)}=-  k_T^{(-)}\psi^{1,(f)}_{(2)}$.
In practice, we extract the scalar amplitudes $\psi(x,\vect{k}_T^2)'s$. 

It is convenient to  classify the $q\bar{q}$ LFWFs by their quark-anti-quark orbital angular momentum (OAM) projected along the $z$-axis, denoted by $l_z$. The angular momentum conservation in $z$-direction then enforces $\Lambda=\lambda+\lambda'+l_z$. Given all possible spin configurations in $\Phi^{\Lambda,(f)}_{\lambda,\lambda'}$, the $l_z$ can be $0$, $\pm 1$ and $\pm 2$, which are s-, p- and d-wave $q\bar{q}$ LFWFs respectively. One can also read off the $l_z$ from the power of $k_T^{(\pm)}$ in Eqs.~(\ref{eq:phi1},\ref{eq:phi2}) directly. Note that in principle all five amplitudes $\psi(x,\vect{k}_T^2)$'s are nonzero, yet in a model calculation or at leading order QED, some of them can be vanishing, which will be shown later. 

\section{The Contact interaction model and photon LFWFs at low virtuality \label{sec:CI}}
\begin{figure}[htbp]       
\includegraphics[width=0.48\textwidth]{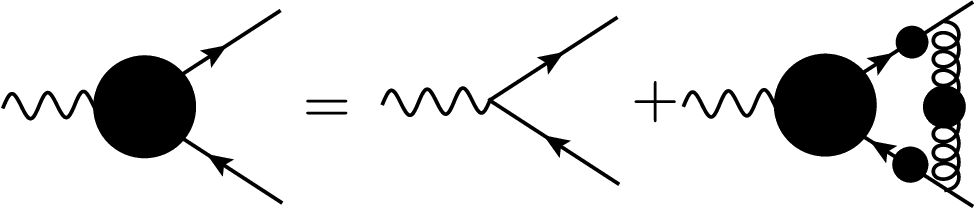}   
 \caption{The diagrammatic representation of inhomogeneous Bethe-Salpeter equation for quark-photon vertex in rainbow-ladder truncation. The black blobs indicate the objects are fully dressed. The dressed gluon propagator here is modeled by contact interaction Eq.~(\ref{eq:Dmunu}).}
\label{fig:photonBSE}           
\end{figure}

The contact interaction model is a simplified model for strong interaction within the Dyson-Schwinger equations approach. Here we recapitulate the formalism based on \cite{Roberts:2010rn}. 
The quark's Dyson-Schwinger equation (or gap equation) formulated in Euclidean space reads
\begin{eqnarray}
\nonumber \lefteqn{S_f(k)^{-1}= i\gamma\cdot k + m_f}\\
&&\hspace{10mm}+ \int \! \frac{d^4q}{(2\pi)^4} g^2 D_{\mu\nu}(k-q) \frac{\lambda^a}{2}\gamma_\mu S_f(q) \frac{\lambda^a}{2}\Gamma_\nu(q;k).
\label{gendse}
\end{eqnarray}
In the contact interaction model, one defines 
\begin{align}
g^2 D_{\mu\nu}(k-q)=\delta_{\mu\nu}\frac{4\pi \alpha_{\rm IR}}{m_G^2} \label{eq:Dmunu}
\end{align}
with $m_G$ a dynamical mass-scale associated with gluon's infrared behavior \footnote{This definition takes the notation used in more recent paper such as \cite{Yin:2019bxe}, which makes a simple replacement $\frac{1}{m_G^2}\rightarrow \frac{4\pi \alpha_{\textrm{IR}}}{m_G^2}$ in \cite{Roberts:2010rn}}. The $\Gamma_\nu(q,k)$ is the Dirac structure part of quark-gluon vertex. Taking the rainbow truncation, i.e., $\Gamma_\nu(q,k)=\gamma_\mu$, and analogously the ladder approximation for quark-antiquark interaction kernel, one arrives at the quark gap equation and quark-photon vertex inhomogeneous BS equation
\begin{align}
 S_f^{-1}(k)& =  i \gamma \cdot k + m_f +  \frac{4}{3} \frac{4\pi\alpha_{\textrm{IR}}}{m_G^2} \int\!\frac{d^4 q}{(2\pi)^4} \,
\gamma_{\mu} \, S_f(q) \, \gamma_{\mu}\,,  \label{eq:gap} \\
\Gamma^{\gamma^*,(f)}_\mu(k;Q)& = \gamma_\mu -  \frac{4}{3} \frac{4\pi\alpha_{\textrm{IR}}}{m_G^2} \int \frac{d^4 q}{(2\pi)^4} \nonumber \\
& \hspace{8mm}\times \gamma_\alpha S_f(q)\Gamma^{\gamma^*,(f)}_\mu(q;Q)S_f(q-Q) \gamma_\alpha\,, \label{eq:gamma}
\end{align}
The Fig.~\ref{fig:photonBSE} displays the Feynman diagram representation for Eq.~(\ref{eq:gamma}). Intuitively, if perturbation theory is applicable, one can see the $\Gamma^{\gamma^*,(f)}_\mu(k;Q)$ is a sum of infinite diagrams containing ladders of one-gluon exchange at all orders. Yet  Eq.~(\ref{eq:gamma}) is essentially nonperturbative and the $\Gamma^{\gamma^*,(f)}_\mu(k;Q)$ contains all the nonperturbative dynamics. Note that the infinite resummation is also encoded in the fully dressed quark propagator $S_f(k)$ as well.

The solution to Eq.(\ref{eq:gap}) is generally
\begin{equation}
\label{eq:genS}
S_f(p)^{-1} = i \gamma\cdot p + M_f\,,
\end{equation}
where $M_f$ is a momentum-independent constant. Meanwhile, the contact interaction kernel eliminates the relative momentum $k$ in $\Gamma^{\gamma^*,(f)}_\mu(k;Q)$ and the general form of the solution to Eq.~(\ref{eq:gamma}) becomes
\begin{align}
\label{eq:GammaQ}
\Gamma^{\gamma^*,(f)}_\mu(Q)= \gamma^T_\mu P^{(f)}_T(Q^2) + \gamma_\mu^L P^{(f)}_L(Q^2)\,, 
\end{align}
where $\gamma^T_\mu = \gamma_\mu-\frac{Q_\mu\slashed{Q}}{Q^2}$ and $\gamma^T_\mu+\gamma^L_\mu=\gamma_\mu$. Note that if the $k-$dependence is kept, there will be ten more Dirac structures \cite{Maris:1999bh}.

In solving Eqs.~(\ref{eq:gap},\ref{eq:gamma}), proper time regularization is employed. The method is to enforce the replacement
\begin{align}
\frac{1}{s+M_f^2} =  \int_0^\infty d\tau\,{\rm e}^{-\tau (s+M_f^2)} \rightarrow  \int_{\tau_{\rm uv}^2}^{\tau_{\rm ir}^2} d\tau\,{\rm e}^{-\tau (s+M_f^2)}
\end{align}
with $\tau_{\rm ir,uv}$ infrared and ultraviolet regulators. Finally, the solution to Eqs.~(\ref{eq:gap},\ref{eq:gamma}) can be summarized as \cite{Roberts:2010rn}
\begin{align}
M_f &= m_f +  \frac{4\alpha_{\rm IR}M_f}{3\pi m_G^2} \,{\cal C}^{\rm iu}(M_f^2)\,,\\
P^{(f)}_L(Q^2)&= 1\,,\\
P^{(f)}_T(Q^2)&= \frac{1}{1+K^{(f)}_\gamma(Q^2)},
\end{align}
with
\begin{align}
K^{(f)}_\gamma(Q^2) =  \frac{4\alpha_{\rm IR}M_f}{3\pi m_G^2}\int_0^1d\alpha\, \alpha(1-\alpha) Q^2\,  \overline{\cal C}^{iu}_1(\omega(M_f^2,\alpha,Q^2))\,
\end{align}
where ${\cal C}^{\rm iu}(M^2)/M^2 = \Gamma(-1,M^2 \tau_{\rm uv}^2) - \Gamma(-1,M^2 \tau_{\rm ir}^2)$, with $\Gamma(\alpha,y)$ being the incomplete gamma-function. The notations $\overline{\cal C}^{iu}_1(z)={\cal C}^{iu}_1(z)/z$, ${\cal C}^{\rm iu}_1(z) = - z (d/dz){\cal C}^{iu}(z)$ and $\omega(M^2,\alpha,P^2) = M^2 + \alpha(1-\alpha) P^2$ are used. The model parameters are set to be $m_G = 0.5$ GeV, $\alpha_{\rm IR}/\pi=0.36$, $\Lambda_{\rm ir} = 0.24$ GeV, $\Lambda_{\rm uv} = 0.91$ GeV, $m_{u/d}=0.007$ GeV and $m_{s}=0.17$ GeV, which well reproduces meson and baryon spectrum \cite{Yin:2019bxe}. These model parameters yield $M_{u/d}=0.37$ GeV and $M_{s}=0.53$ GeV, so $K^{(f)}_\gamma(Q^2)$ can be determined.

We can now calculate the $q\bar{q}$ LFWF of photon through Eq.~(\ref{eq:chi2phi}). We first perform a rearrangement of $\Gamma^{\gamma^*,(f)}_\mu(Q)$ in Eq.~(\ref{eq:GammaQ}) to be 
\begin{align}
\label{eq:GammaQQ}
\Gamma^{\gamma^*,{(f)}}_\mu(Q) = \gamma_\mu P^{(f)}_T(Q^2) + \gamma_\mu^L [P^{(f)}_L(Q^2)-P^{(f)}_T(Q^2) ]\,.
\end{align}
In this way, the first term has the Dirac structure of bare vertex, which is the leading order vertex in QED. Meanwhile, a few notations and relations between four momentums in the Euclidian space are as follows. A four momentum in Euclidian space is $k=(k_1,k_2,k_3,k_4)$, with first three components corresponding to $x$, $y$ and $z$ directions. We separate it into longitudinal and transverse vectors, e.g., $k_\para =(\textbf{0},k_3,k_4)\equiv (\textbf{0},\vect{k}_\para)$ and $k_\perp=(\vect{k}_\perp,\textbf{0})$. In Minkowski space, denoting the four vector of a photon moving in $z-$direction as  $q=(q^0,\textbf{0},q^3)$, which satisfies $q^2=-Q^2$, the polarization vector is taken as $\epsilon_0=(q^3/Q,\textbf{0},q^0/Q)$ \cite{Dosch:1996ss,Kovchegov:2012mbw}, where abbreviation  $Q\equiv\sqrt{Q^2}$ is used. We introduce a light front null vector $n=1/\sqrt{2}(1,0,0,-1)$, which satisfies $n^2=0$ and can be used to project out the plus component of a four vector, i.e., $n\cdot A=A^+$.  
Making use of the relation $(A\cdot B)_M\rightarrow-(A\cdot B)_E$ from Minkowski to Euclidian space, one can obtain the following useful identities in Euclidian space: $k_\perp \cdot n=0$, $\epsilon_0 \cdot n=Q \cdot n/Q$, $\epsilon_0 \cdot k_\perp=0$ and $\epsilon_0 \cdot Q=0$. Here $Q$ denotes the four momentum of photon in Euclidian space, and we will use it and $Q=\sqrt{Q^2}$ without distinction. But they can be easily distinguished from the context. Starting with Eqs.~(\ref{eq:chi2phi},\ref{eq:genS},\ref{eq:GammaQQ}), and omit the flavor index for abbreviation, the derivation goes as
\begin{widetext}
\begin{align}
	\langle x \rangle^m &\equiv\int_0^1 dx x^m \Phi^0_{+,-}(x,\vect{k}_T) \label{xm1}\\
	&=-\frac{e_f e P_T(Q^2)}{2\sqrt{3}}\int \frac{d^2 \vect{k}_{\parallel}}{2 \pi} \left(\frac{k^+}{Q^+}\right )^m\frac{1}{|Q^+|}\textrm{Tr}\Big [(I+\gamma^5)\gamma^+ S(k)[\Gamma^{\gamma^*}(k;Q)\cdot \epsilon_0(Q)]S(k-Q)  \Big ]  \label{xm2}\\
	&=-\frac{e_f e P_T(Q^2)}{2 \sqrt{3}|Q \cdot n|} \int \frac{d^2 \vect{k}_{\parallel}}{2 \pi} \left(\frac{k_\para \cdot n}{Q \cdot n}\right )^m \frac{{\rm Tr}[\slashed{n}(-i \slashed{k}+M)\slashed{\epsilon}_0(-i\slashed{k}+i\slashed{Q}+M)]}{(k^2+M^2)(k^2-2 k \cdot Q+Q^2+M^2)}  \label{xm3}\\
	&=-\frac{\sqrt{N_c}e_f e P_T(Q^2)}{2 |Q \cdot n|} \int \frac{d^2 \vect{k}_{\parallel}}{2 \pi} \left(\frac{k_\para \cdot n}{Q \cdot n}\right )^m \frac{-2\sqrt{2}[(k_\para^2+k_\perp^2+M^2-Q\cdot k_\para)(-\sqrt{2}\epsilon_0 \cdot n)+\sqrt{2}(2 n\cdot k_\para-Q\cdot n)(\epsilon_0 \cdot k_\para)]}{(k_\para^2+k_\perp^2+M^2)(k_\para^2-2 k_\para \cdot Q+Q^2+M^2+k_\perp^2)} \label{xm4}\\
	&=\frac{\sqrt{2N_c}e_f e P_T(Q^2)}{|Q \cdot n|}\int_0^1 du\int \frac{d^2 \vect{k}_{\parallel}}{2\pi} \left(\frac{k_\para \cdot n}{Q \cdot n}\right )^m \frac{(k_\para^2+k_\perp^2+M^2-Q\cdot k_\para)(-\sqrt{2}\epsilon_0 \cdot n)+\sqrt{2}(2 n\cdot k_\para-Q\cdot n)(\epsilon_0 \cdot k_\para)}{[u(k_\para^2+k_\perp^2+M^2)+(1-u)(k_\para^2-2 k_\para \cdot Q+Q^2+M^2+k_\perp^2)]^2} \label{xm5}\\
	&=\frac{\sqrt{2N_c}e_f e P_T(Q^2)}{|Q \cdot n|}\int_0^1 du\int \frac{d^2 \vect{k}_{\parallel}}{2\pi} \left(\frac{k_\para \cdot n}{Q \cdot n}\right )^m \frac{(k_\para^2+k_\perp^2+M^2-Q\cdot k_\para)(-\sqrt{2}\epsilon_0 \cdot n)+\sqrt{2}(2 n\cdot k_\para-Q\cdot n)(\epsilon_0 \cdot k_\para)}{[(k_\para-(1-u)Q)^2+\Delta^2]^2} \label{xm6}\\
	&=\frac{\sqrt{2N_c}e_f e P_T(Q^2)}{|Q \cdot n|}\int_0^1 du \int \frac{d^2 \vect{k}_{\parallel}}{2\pi}\frac{[k_\perp^2+M^2-(1-u)u Q^2](1-u)^m(-\sqrt{2}\epsilon_0 \cdot n)}{(k_\para^2+\Delta^2)^2} \label{xm7} \\
	&=\frac{2\sqrt{N_c}e_f e P_T(Q^2)}{Q}\int_0^1 du' u'^m \int\frac{d^2 \vect{k}_{\parallel}}{2\pi}\frac{k_\perp^2+M^2-u'(1-u')Q^2}{[k_\para^2+Q^2 u'(1-u')+M^2+k_\perp^2]^2} \label{xm8}\\
	&=\int_0^1 du' u'^m\frac{\sqrt{N_c}e_f e P_T(Q^2)}{Q}\left(1-\frac{2u'(1-u')Q^2}{Q^2 u'(1-u')+M^2+k_\perp^2}\right). \label{xm9}
\end{align}
\end{widetext}
Comparing Eq.~(\ref{xm1}) and Eq.~(\ref{xm9}), we deduce
\begin{align}
\Phi_{+,-}^0(x,\vect{k}_T)=e_f e P_T(Q^2)\frac{\sqrt{N_c}}{Q}\left(1-\frac{2x(1-x)Q^2}{k_\perp^2+Q^2 x(1-x)+M^2}\right). \label{eq:phi01}
\end{align}
From Eq.~(\ref{xm2}) to Eq.(\ref{xm3}), the trace operation in color space produces an overall color factor $N_c$ and unity in flavor space. Feynman parametrization technique is implemented in getting Eq.~(\ref{xm5}). From Eq.~(\ref{xm6}) to Eq.~(\ref{xm7}), we have performed a momentum shift in $\vect{k}_\para$. The numerator is fully expanded into polynomials and due to $n^2=0$, only a few terms survive after integration over $\vect{k}_\para$. We also changed the variable $u$ to $1-u'$ in getting Eq.~(\ref{xm8}), and the identity 
\begin{align}
\int d^2 k\frac{1}{(k^2+A^2)^2}=\frac{\pi}{A^2} 
\end{align}
is used in getting Eq.~(\ref{xm9}). Finally, we remark that employing the proper-time regularization scheme in calculating the momentum integral in Eq.~(\ref{xm8}) might be more appropriate from the perspective of effective model. Yet it would hinder a direct comparison with LO QED LFWFs, and since the integral converges, we do not impose the proper-time regularization.

\begin{figure}[htbp]       
\includegraphics[width=0.48\textwidth]{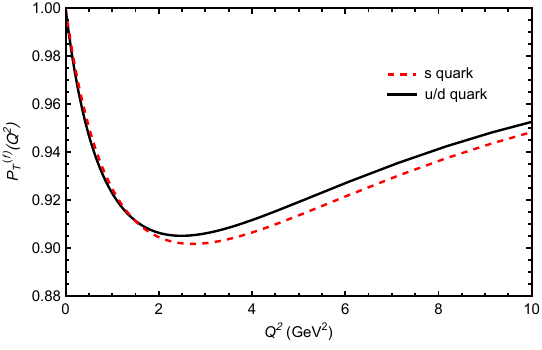}   
 \caption{The $P_T^{(u/d)}(Q^2)$ and $P_T^{(s)}(Q^2)$.}
\label{fig:PT}           
\end{figure}   

Analogously, using Euclidian space relations $\epsilon_{\pm 1}\cdot Q=0$, $\epsilon_{\pm 1}\cdot n=0$, $\epsilon_{\pm 1}\cdot k_\perp=\mp k_T^{(\pm)}$ and $\epsilon_{\pm 1}\cdot k_\para=0$ in photon's collinear reference frame ($\vect{Q}_\perp=\textbf{0}$), we can derive all other photon $q\bar{q}$ LFWFs, which is summarized as
\begin{align}
\psi_{(1)}^{0,(f)}(x,\vect{k}_T^2)&\xlongequal[]{\textrm{reduced}}-e_f e P^{(f)}_T(Q^2)\frac{\sqrt{N_c}}{Q}\frac{2x(1-x)Q^2}{k_\perp^2+Q^2 x(1-x)+M_f^2}, \label{eq:psi01} \\
\psi^{0,(f)}_{(2)}(x,\vect{k}_T^2)&=0,\\
\psi^{1,(f)}_{(1)}(x,\vect{k}_T^2)&=e_f e P^{(f)}_T(Q^2)\sqrt{2N_c}\frac{M_f}{k_\perp^2+Q^2 x(1-x)+M_f^2},\\
\psi^{1,(f)}_{(2)}(x,\vect{k}_T^2)&=e_f e P^{(f)}_T(Q^2)\sqrt{2N_c}\frac{x}{k_\perp^2+Q^2 x(1-x)+M_f^2},\\
&=-\psi^{1,(f)}_{(3)}(1-x,\vect{k}_T^2),\\
\psi^{1,(f)}_{(4)}(x,\vect{k}_T^2)&=0. \label{eq:psi14}
\end{align}
Note that in rewriting  Eq.~(\ref{eq:phi01}) into Eq.~(\ref{eq:psi01}), we have omitted the unity term, as it becomes a Dirac delta function after Fourier transform with respect to $\vect{k}_T$ and does not contribute in color dipole model regarding cross sections we are interested in. 

There are a few remarks worth addressing. Since $\epsilon_\Lambda \cdot Q=0$, which leads to $\epsilon_\Lambda \cdot \gamma_L=0$, the quark photon vertex in Eq.~(\ref{eq:GammaQQ}) has only one Dirac structure $\gamma_\mu$ that contributes to Eq.~(\ref{eq:chi2phi}). So the quark-photon vertex is equivalent to bare vertex $\gamma_\mu$ and the quark propagator takes a constituent quark mass $M$. These are exactly the same structures used in \cite{Dosch:1996ss} to obtain photon LFWF, only that therein the light-cone perturbation method is employed. In this sense, our derivation confirms that both approaches yield analytically same result. Yet our approach doesn't rely on perturbation method, and is nonperturbative in essence. In Eqs.~(\ref{eq:psi01}-\ref{eq:psi14}), the intrinsic nonperturbative information is encoded in the constituent quark mass $M$ and the quark-photon vertex dressing factor $P_T(Q^2)$. We remind that the $M_{u/d/s}$ is generally two orders of magnitude larger than the current quark mass. It is a direct reflection of the chiral symmetry breaking in QCD, as well as the nonperturbative dynamics. In this connection, a quark mass around 140 MeV that was popular in the CGC model fitting to the HERA data \cite{Watt:2007nr} seems to be a compromise between constituent quark mass around $300-400$ MeV and parton-like current quark mass. In addition, we also notice a mulitiplicative (and nonperturbative) factor $P_T(Q^2)$ in Eq.~(\ref{eq:phi01}). It brings a few percents of suppression to photon $q\bar{q}$ LFWFs at low $Q^2$, as shown in Fig.~\ref{fig:PT}, but reduces to unity at $Q^2=0$.

The limitations in our result should also be reminded. Firstly, due to the contact interaction model, the quark-photon vertex has only one Dirac structure $\gamma_\mu$ that contributes. If one employs more realistic model that allows  $\Gamma^{\gamma^*}_\mu(k;Q)$ to be dependent on $k$, 10 more Dirac structures would appear,  such as $k_\mu, k_\mu \slashed{Q}$ and etc. In that case, the $\psi_{(2)}^0$ and $\psi_{(4)}^1$ will no longer be vanishing, i.e., p- and d-wave components could appear. Such property has been observed in the study of vector meson LF-LFWFs \cite{Shi:2021taf}. Secondly, the contact interaction model works at low energy so our result Eqs.~(\ref{eq:psi01}-\ref{eq:psi14}) only hold for low $Q^2$. At large $Q^2$, the photon LFWFs should approach the perturbative result. Therefore we expect a complete photon LFWF from low to high $Q^2$ should undergo a transition from our nonperturbative LFWFs to the perturbative ones. 

\section{Incorporating nonperturbative photon $q\bar{q}$ LFWFs in small-$x$ DIS. \label{sec:DIS}}
The inclusive $ep$ DIS at small $x$ can be described by color dipole model \cite{Nikolaev:1990ja,Mueller:1993rr,Kowalski:2003hm}, in which a virtual photon is emitted by the incoming electron, and then splits into a color dipole that scatters with proton inelastically.  The cross section takes the factorized form 
\begin{align}
\sigma_{L,T}^{\gamma^*p}(x,Q^2)&=2 \sum_{f}\int d^2\vect{r}\int d^2\vect{b}\int_0^1dz|\Psi^{(f)}_{L,T}(z,r;Q^2)|^2  \nonumber\\ 
&\ \ \ \times {\cal N}(x^{(f)},r,b),\label{eq:sigmarLT}
\end{align}
where 
\begin{align}
|\Psi^{(f)}_{T}(r,z;Q^2)|^2&=\frac{1}{2}\sum_{\lambda,\bar{\lambda'}}\left[ |\tilde{\Phi}_{\lambda,\bar{\lambda'}}^{1,(f)}(r,z;Q^2)|^2+\tilde{\Phi}_{\lambda,\bar{\lambda'}}^{-1,(f)}(r,z;Q^2)|^2 \right],\\
|\Psi^{(f)}_{L}(r,z;Q^2)|^2&=\sum_{\lambda,\bar{\lambda'}}|\tilde{\Phi}_{\lambda,\bar{\lambda'}}^{0,(f)}(r,z;Q^2)|^2,
\end{align}
are the squared photon $q\bar{q}$ LFWFs. The $\tilde{\Phi}$ is defined as the photon $q\bar{q}$ LFWFs in coordinate space, e.g.,
\begin{align}
\tilde{\Phi}^{\Lambda,(f)}_{\lambda,\lambda'}(z,\vect{r})=\int \frac{d^2\vect{k}}{(2\pi)^2}\textrm{e}^{i\vect{k}\cdot\vect{r}} \Phi^{\Lambda,(f)}_{\lambda,\lambda'}(z,\vect{k}).
\end{align}
In this work we take the light quarks (u,d and s) and charm quark into account. In this case the $x^{(f)}$ equals Bjorken-$x$ for light quarks, and $x^{(c)}=x(1+4m_c^2/Q^2)$ for charm quark.  

The ${\cal N}(x,r,b)$ is the imaginary part of the
forward dipole-proton scattering amplitude, with color dipole
transverse size $r$ and collision impact parameter $b$. Here we employ the bCGC model \cite{Watt:2007nr,Rezaeian:2013tka}, which reads 
\begin{eqnarray}
 {\cal N}(x,r,b)=\left\{
 \begin{aligned}
& N_0\left(\frac{r Q_s}{2}\right)^{2\gamma_{\rm eff}} \hspace{20 mm} rQ_s\le 2, \\ 
& 1-{\rm exp}[-{\cal A} {\rm ln}^2({\cal B}rQ_s)]  \hspace{5.2 mm} rQ_s >2,
 \end{aligned}
 \right.
\end{eqnarray}
with 
\begin{align}
Q_s(x,b)&=\left(\frac{x_0}{x}\right)^{\frac{\lambda}{2}} {\rm exp}\left[-\frac{b^2}{4\gamma_s B_{\rm CGC}}\right], \\
\gamma_{\rm eff}&=\gamma_s+\frac{1}{\kappa \lambda Y}{\rm ln}\left(\frac{2}{r Q_s}\right),  \\
Y&={\rm ln}(1/x),
\end{align}
and 
\begin{align}
{\cal A}&=-\frac{N_0^2 \gamma_s^2}{(1-N_0)^2{\rm ln}(1-N_0)},\\
{\cal B}&=\frac{1}{2}(1-N_0)^{-\frac{1-N_0}{N_0 \gamma_s}}.
\end{align}
The $\kappa=9.9$ is fixed to be LO BFKL value, and $B_{\rm CGC}=5.5$ GeV$^{-2}$ determined by fitting exclusive meson production data in \cite{Rezaeian:2013tka}. Other model parameters  $N_0$, $\gamma_s$, $x_0$ and $\lambda$ are determined by fitting inclusive DIS data \cite{Rezaeian:2013tka}.  In this work, we employ the reduced cross section $\sigma_r$ data from HERA \cite{H1:2015ubc}, which is a linear combination of proton structure functions $F_2$ and $F_L$, i.e.,
\begin{align}
\sigma_r(x,y,Q^2)=F_2(x,Q^2)-\frac{y^2}{1+(1-y)^2}F_L(x,Q^2)
\end{align}
with 
\begin{align}
F_2(x,Q^2)&=\frac{Q^2}{4\pi^2\alpha_{\rm em}}[\sigma_L^{\gamma^*p}(x,Q^2)+\sigma_T^{\gamma^*p}(x,Q^2)], \\
F_L(x,Q^2)&=\frac{Q^2}{4\pi^2\alpha_{\rm em}}\sigma_L^{\gamma^*p}(x,Q^2).
\end{align}
The $y=Q^2/(s x)$ is the inelasticity parameter and $s$ is the center of mass energy square. 

From Eq.~(\ref{eq:sigmarLT}) we can see the inclusive DIS cross section is the convolution between the square of LFWFs and color dipole model. This makes the $\sigma_r(x,y,Q^2)$ sensitive to photon LFWFs. From high to low $Q^2$, the color dipole gets larger in size and nonperturbative effect should emerge. We can therefore investigate whether the  nonperturbative photon $q\bar{q}$ LFWFs can play a role at low $Q^2$. 

As perturbative photon $q\bar{q}$ LFWFs are applicable at large $Q^2$, and our nonperturbative photon $q\bar{q}$ LFWFs only work at low $Q^2$, we introduce a modified model that linearly combines the two, i.e.,
\begin{align}\label{eq:psiprime}
|\Psi'^{(f)}_{T,L}(r,z;Q^2)|^2&=F_{\rm part}(Q^2)|\Psi^{(f),{\rm np}}_{T,L}(r,z;Q^2)|^2 \nonumber \\
&\ \ \ \ +[1-F_{\rm part}(Q^2)]|\Psi^{(f),{\rm p}}_{T,L}(r,z;Q^2)|^2
\end{align}
with a $Q^2-$dependent weighting factor 
\begin{align}
F_{\rm part}(Q^2)=\frac{Q_0^{2n}}{(Q^2+Q_0^2)^n}.\label{eq:Fpart}
\end{align}
We remind that the nonperturbative $\Psi^{(f),\textrm{np}}_{T,L}$ uses Eqs.~(\ref{eq:psi01}-\ref{eq:psi14}), and the perturbative term $\Psi^{(f),\textrm{p}}_{T,L}$ can be obtained by setting $P_T(Q^2)=1$ and $M_f=m_f$ in Eqs.~(\ref{eq:psi01}-\ref{eq:psi14}). The two parameters $Q_0$ and $n$ modulate the rising of nonperturbative effect. We further assume the $Q_0$ and $n$ are the same for $u/d$ and $s$ quarks. For the heavy charm quark, we ignore the nonperturbative effect for now, as it is supposed to be small due to the large intrinsic scale brought by quark mass. A quantitative check will be left for future investigation.
At large $Q^2$, $F_{\rm part}(Q^2)$ approaches zero so the $|\Psi'^{(f)}_{T,L}(r,z;Q^2)|^2$ is dominated by perturbative photon $q\bar{q}$ LFWFs $|\Psi^{(f),{\rm p}}_{T,L}(r,z;Q^2)|^2$. As $Q^2$ decreases, $F_{\rm part}(Q^2)$ increases and $|\Psi'^{(f)}_{T,L}(r,z;Q^2)|^2$ starts to pick up contributions from nonperturbative photon LFWFs $|\Psi^{(f),{\rm np}}_{T,L}(r,z;Q^2)|^2$, until the LFWFs become completely nonperturbative at $Q^2=0$ GeV$^2$.  

\begin{table*}[t!]
  \centering
 \tabcolsep=0.25cm
  \begin{tabular}{ccccccccc}
    \hline\hline
LFWFs [Eqs.~(\ref{eq:psi01}-\ref{eq:psi14},\ref{eq:psiprime})]&  $Q^2/\text{GeV}^2$ &  $\gamma_s$  &$N_0$ & $x_0$ & $\lambda$   &$Q_0^2$ & $n$ & $\chi^2$/d.o.f \\ \hline
Pert.  &  $[0.85,50]$  &  $ 0.6290 $ & $0.4199$ & $2.395 \times 10^{-4}$ &$0.1962$&-&-&$265.8/223=1.192$ \\
Pert.  &  $[0.25,50]$  &  $ 0.3869 $ & $0.7556$ & $7.047 \times 10^{-7}$ &$0.1052$&-&-&$678.4/282=2.406$ \\
Pert.+Nonpert. & $[0.25,50]$  & $ 0.6177 $ & $0.4596$ & $1.326 \times 10^{-4}$ &$0.1875$&$1.052$&$3.970$&$337.9/280=1.207$ \\
 \hline
  \end{tabular}
  \caption{Fitting the reduced DIS cross section data of HERA \cite{H1:2015ubc} with bCGC model using nonperturbative-effect modified photon LFWFs (Eq.~\ref{eq:psiprime}). The current quark masses are set to physical values $m_{u/d}=4$ MeV, $m_s=95$ MeV and $m_c=1.27$ GeV. The dressed quark mass $M_{u/d/s}$ and dressing function $P^{(u/d/s)}_T(Q^2)$ are determined by contact interaction model in Sec.~\ref{sec:CI}.}
  \label{tab:1}
\end{table*}

\begin{figure}[htbp]       
\includegraphics[width=0.5\textwidth]{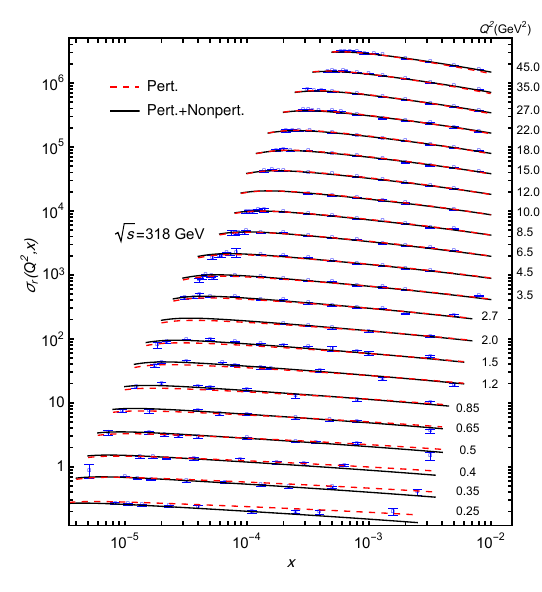}   
 \caption{The reduced cross section $\sigma_r$ of DIS at $\sqrt{s}=318$ GeV$^2$. Data points are taken from \cite{H1:2015ubc} and curves are calculated using parameters from the third (black solid) and second (red dashed) rows of Table.~\ref{tab:1}.}
\label{fig:sigmar318}           
\end{figure}   

\begin{figure}[htbp]       
\includegraphics[width=0.5\textwidth]{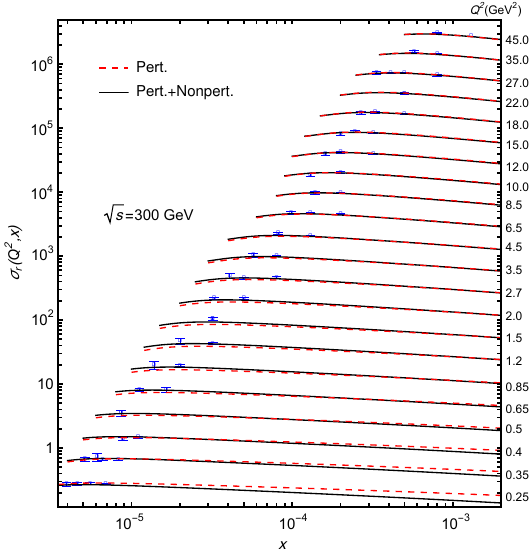}   
 \caption{The reduced cross section $\sigma_r$ of DIS at $\sqrt{s}=300$ GeV$^2$. Data points are taken from \cite{H1:2015ubc} and curves are calculated using parameters from the third (black solid) and second (red dashed) rows of Table.~\ref{tab:1}.}
\label{fig:sigmar300}           
\end{figure}   

We then fit the reduced cross section data \cite{H1:2015ubc} with $|\Psi'^{(f)}_{T,L}(r,z;Q^2)|^2$. The employed data is limited to $x<0.01$ and $Q^2<50$ GeV$^2$. We remind that in \cite{Rezaeian:2013tka}, the authors found that when current quark masses are employed in perturbative photon LFWFs, the bCGC model can well fit HERA data only for $Q^2\ge 0.85$ GeV$^2$. This is reasonable, as the photon LFWFs can only approach the perturbative form for certain large $Q^2$. Here with nonperturbative corrections, our result is shown in Table.~\ref{tab:1}. The first column indicates whether perturbative (LO QED and QCD) or the nonperturbative-effect modified photon LFWFs are employed. The first row shows that data of $Q^2 \in [0.85,50]$ GeV$^2$ can be fitted using perturbative photon LFWFs with current quark masses, in agreement with \cite{Rezaeian:2013tka}. Here the determined parameters are a bit different from those in \cite{Rezaeian:2013tka} as we use $m_{u/d}=4$ MeV and $m_s=95$ MeV, while therein $m_s=m_{u/d}\approx 0$ MeV.
 To investigate nonperturbative effects, we then employ data to lower $Q^2$, e.g., $Q^2\approx \Lambda_{\textrm{QCD}}^2$. We find that for $Q^2 \in [0.25,50]$ GeV$^2$, the fitting with perturbative photon LFWFs (second row) becomes significantly worse, while that incorporating nonperturbative effect (third row) remains good \footnote{ Note that to accommodate low $Q^2$ data, the bCGC model parameters in the third row are quite different from those in the first row. However, if the model parameters in row one are used, we will get $\chi^2\approx 2000$ for data of $Q^2 \in [0.25,50]$ GeV$^2$, which is unreasonable.}.
Figs.~\ref{fig:sigmar318} and \ref{fig:sigmar300} show the HERA $\sigma_r$ data as compared to calculated cross section using parameters in the second (red dashed) and third (black solid) rows of Table.~\ref{tab:1}. Since $F_{\rm part}(Q^2)$ is a positive definite function, this result strongly suggests that incorporating the nonperturbative photon $q\bar{q}$ LFWFs can significantly improve the agreement between color dipole model calculations and inclusive small-$x$ DIS data at low $Q^2$.

\section{Summary\label{sec:sum}}
Circuvmenting the light front quantization procedure, the photon $q\bar{q}$ LFWFs can be extracted from covariant un-amputated quark-photon vertex, which is available from various nonperturbative QCD methods or models that are quantized in ordinary space-time. In this paper, we resort to a contact interaction model. With this particular model, the nonperturbative effects are encoded in the enhanced quark mass $M$ and a dressing factor $P_T(Q^2)$ that both arise through nonperturbative dynamics, as compared to LO QED photon LFWFs.  In particular, for a real photon $P_T(Q^2=0)=1$, so the nonperturbative effects are totally absorbed into the enhanced quark mass $M$. The QCD's dynamical chiral symmetry breaking (DCSB) property thus plays a significant role in shaping the photon $q\bar{q}$ LFWFs at low $Q^2$.

Within color dipole model, the cross section of inclusive DIS is an integral of the squared photon $q\bar{q}$ LFWFs, hence sensitive to photon LFWFs. We then modify the perturbative photon LFWFs by incorporating nonperturbative effects at low $Q^2$ through Eq.~(\ref{eq:psiprime}). It is found that this modification can significantly improve the agreement between color dipole model and small-$x$ inclusive DIS HERA data toward lower $Q^2$ region, in agreement with earlier works using a phenomenological model \cite{Forshaw:2003ki,Flensburg:2008ag,Berger:2012wx,Goncalves:2020cir}.

Finally, we remind that in present work, the nonpertubrative photon LFWFs in Eqs.~(\ref{eq:psi01}-\ref{eq:psi14}) are limited to some unknown low $Q^2$. They do not transform into perturbative ones at large $Q^2$ by themselves. The transition is thus modeled by a primitive function Eq.~(\ref{eq:Fpart}). Yet this limitation is not brought by the projection method Eq.~(\ref{eq:chi2phi}), but roots in the contact interaction model, which only works at low scales. It can be overcame by employing covariant unamputated quark-photon vertex that contains both infrared and ultra-violate dynamics. In this connection, the Maris-Tandy model within Dyson-Schwinger equations approach \cite{Maris:1997tm,Maris:1999nt}, along with other studies on quark-photon vertex formulated in ordinary space time, can be considered as good candidates for future study.

\begin{acknowledgments}
Chao Shi is grateful for Ya-Ping Xie and Pei-Lin Yin for helpful suggestions. This work was supported by the National Natural Science Foundation of China (under Grant No. 11905104, No. 12165004); the Education Department of Guizhou Province under Grant No.QJJ[2022]016; and the Strategic Priority Research Program of Chinese Academy of Sciences (Grant NO. XDB34030301).
\vspace{3em}
\end{acknowledgments}

\appendix

\bibliography{PhotonLFWF}

\end{document}